# Optimization of vertically mounted agrivoltaic systems

Pietro Elia Campana[1*], Bengt Stridh[1], Stefano Amaducci[2], Michele Colauzzi[2]

[1]Mälardalen University, Future Energy Center, Box 883, Västerås, Sweden

[2]Università Cattolica del Sacro Cuore, Department of Sustainable Crop Production, via Emilia Parmense 84, Piacenza, Italy

*Corresponding author (pietro.campana@mdh.se)

## Abstract

Agrivoltaic systems represent a key technology for reaching sustainable development goals reducing the completion of land for food versus land for energy. Moreover, agrivoltaic systems are at the centre of the nexus between electricity production, crop production, and irrigation water saving. In this study, an optimization model for vertically mounted agrivoltaic systems with bifacial photovoltaic models is developed. The model combines three main submodels: solar radiation and shadings, photovoltaic, and crop yield. Validation of the submodels is performed showing good agreement with measured data and commercial software. The optimization model is set as multi objective to explore the trade-offs between competing agrivoltaic key performance indicators.

The results shows that the row distance between bifacial photovoltaic modules structure affects significantly the photosyntetically active radiation distribution by reducing the crop yield of potato and oat of about 50% by passing from 20 m to 5 m. The implementation of agrivoltaic system for the investigated crops at the chosen location shows a land equivalent ratio above 1.2 that justify the technology for reaching the country sustainability goals.

## 1  Introduction

At the end of 2019, the photovoltaic system capacity in Sweden was 698 MW$_p$ with a total estimated production of about 540 GWh that accounted for 0.41% of the total electricity supply (Lindahl et al., 2020). At the end of 2019, ground-mounted centralized PV parks represented only the 5% of the total grid connected market in Sweden, nevertheless, as highlighted by Lindahl et al. (2020), this market segment has significantly increased in the last two years and it is expected to increase in the next years. This is also boosted by the Swedish Government targets of 100% purely renewable-based power supply for 2040 (Government Offices of Sweden, 2021a) and net zero emissions of greenhouse gases into the atmosphere by 2045 (Government Offices of Sweden, 2021b). Despite the rapid development of the PV sector in Sweden, the PV installations still suffer of a major issue, the system profitability. The profitability is threatened by 1) low electricity prices, 2) low annual solar irradiation, 3) seasonality of the solar radiation, and 4) lower subsidies as compared to other southern European countries like Italy (D´Adamo et al., 2020; D´Adamo et al., 2021). Those aspects have been clearly discussed in a recently published study on PV systems optimization that covers the entire Sweden (Campana et al., 2020). New solutions and business models are required to make solar PV systems and in particular solar parks more economically profitable to easily pursue the national targets of high renewable penetrations and net zero emissions. Traditionally, land based PV farms have increased the competition of land resources for food production (Nonhebel, 2005). Nevertheless, in recent years, researchers mostly from France, Germany,

Italy, and USA have proved that agrivoltaic systems, that are the combination of farm activities and PV farms, can overcome the competition land for food vs land for energy.

Agrivoltaic systems presents several advantages over traditional ground-based PV systems and by adopting a holistic approach, those advantages are crosscutting among three different macro-areas: energy, food, and water. First, the combination of power and crop production can increase the economic benefits of the entire system. By combining synergies and revenues, the payback period of PV systems can decrease making ground-based PV systems more attractive and thus easily supports the national renewable energy targets. From a pure energy perspective, agrivoltaic systems can increase the electricity production of solar panels due to the microclimate (i.e., lower operating temperatures on the back side of the solar panels and thus higher efficiency) created by the growing crops and higher PV modules installations height as compared to ground based PV systems. Moreover, by combining farm activities, such as for instance irrigation, and PV production, the self-consumption rate is higher since most of the agricultural activities are concentrated during those months with higher solar radiation and thus production. Higher self-consumption rates can lead to several benefits into the electric grid by avoiding congestion and thus increasing revenues due to the mismatch between selling and buying electricity prices. From a food and water perspective, as highlighted in the background, agrivoltaic systems have the potentials lead to higher crop yields because the shadowing from optimally distributed PV panels decreases the amount of water required by the crops and increases the soil moisture. This is of paramount importance, especially in the context of the negative effects that climate changes and weather extremes (i.e., increased temperatures, decreased precipitation, and drought) can have on the crop yields.

A typical configuration of an agrivoltaic system consists in having the PV modules installed at a height of 2-5 meters above ground, thanks to suspended structures, to allow normal farm activities underneath. This concept was firstly introduced in the 1980s from Goetzberger and Zastrow (1982). Nevertheless, one of the first agrivoltaic experiment was conducted in France in 2013 by Marrou et al. (2013) using vegetables as crop. The authors conducted an experiment with two shading levels equal to 50% and 70% and observed a negligible change in the Radiation Conversion Efficiency. The authors suggested that optimization was required to find the best trade off of PV modules densities for balancing food production and electricity production. In a previous work conducted by the same research group, it was proved through modelling that the combination of PV modules and agriculture could lead to an increased land productivity of 60-70% (Dupraz et al., 2011). More recently, Dinesh and Pearce (2016) have shown that agrivoltaic system can increase the farms´ generated economic value of about 30% as compared to common practices in the USA. In 2018, while a severer drought affected almost the entire Europe, the Fraunhofer Institute for Solar Energy Systems (2021) obtained a land use efficiency of about 180% as compared to the reference case of only crop production (100% efficiency) or electricity production (100% efficiency) in the experiment carried out in Heggelbach. Amaducci et al. (2018) demonstrated that agrivoltaic systems with an optimal density of PV modules can also increase crop yields since they can keep higher level of soil moisture. The study is based on an extensive simulation approach of agrivoltaic system in Northern Italy using 40 years climatological data, several agrivoltaic system configurations, and scenarios. The work carried out by Amaducci et al. (2018) is considered a milestone for the technology since it indicated that agrivoltaic systems can support crop yield, clean energy production, and water saving, playing thus a key role in the energy-food-water nexus. Moreover, the study showed that agrivoltaic systems can improve crop resilience to climate change and weather induced phenomena like droughts. No optimization was set up and performed in Amaducci et al. (2018). More recently, Trommsdorff et al. (2021) studied through simulations and sensitivity analysis the effect of an agrivoltaic system´s design parameters on the solar radiation availability on crops grown underneath of the agrivoltaic system. In all the cases and scenarios investigated, the land equivalent ratio was higher than one (i.e., the land productivity, considering both electricity and food production, is higher with agrivoltaic systems than when the same land is employed for mono production). Barron-Gafford et al. (2019) also studied the energy-water-food nexus of agrivoltaic systems. Their results showed that the combination of PV systems with farm activities could lead to reduced plant drought stress, and higher food yield, similar to Amaducci et al. (2018). Nevertheless, for the first time, through a fully monitored experiment, the authors showed that agrivoltaic systems can reduce PV panels heat stress leading to highest energy conversion efficiencies. Economic aspects of agrivoltaic system were studied in Schindele et al. (2020), and in Agostini et al. (2021), nevertheless those studies were mostly economic assessments of specific existing agrivoltaic systems and optimization was not included.

Agrivoltaic systems are fully embracing the concept of green economy by promoting low carbon technologies, reducing pollution, using resources efficiently, boosting employment, and preventing loss of agroecosystem services (UNEP, 2021). Agrivoltaic systems represents also a key technology to meet the sustainable development goals especially, zero hunger, clean and affordable energy, and climate action (United Nations, 2021). The technology itself promote the integration and the investigation of multidisciplinary areas such as energy, food, water, environment, and ecology that are essential for reaching the sustainable development targets (Baleta et al., 2019).

To the best knowledge of the authors, there is a research gap in the field of agrivoltaic related to the system techno-economic optimization. Moreover, to the best knowledge of the authors, extremely few research activities have been conducted on agrivoltaic systems in Sweden. Most of the research efforts have been focused on solar irrigation systems (Campana et al., 2015) and integration of PV systems and greenhouses (Vadiee, 2013). From the report carried out by Lindahl et al. (2020), most of the application of PV systems in agriculture are simply related to the integration of PV systems on farms' buildings. This is mostly due to the interdisciplinarity of the field, with some of the researchers focusing on crop modelling and other researchers focusing on PV system modelling. The aim of this study is to develop a techno-economic optimization model for agrivoltaic systems. This is done by leveraging on our previous study on water-food-energy nexus (Campana et al., 2018) and PV systems modelling (Campana et al., 2018). A techno-economic optimization model is developed for the optimization of design parameters of vertically mounted agrivoltaic system using bifacial PV models as shown in Figure 1. This agrivoltaic system configuration has been chosen since it is the same technology used for the implementation of the first agrivoltaic system in Sweden.

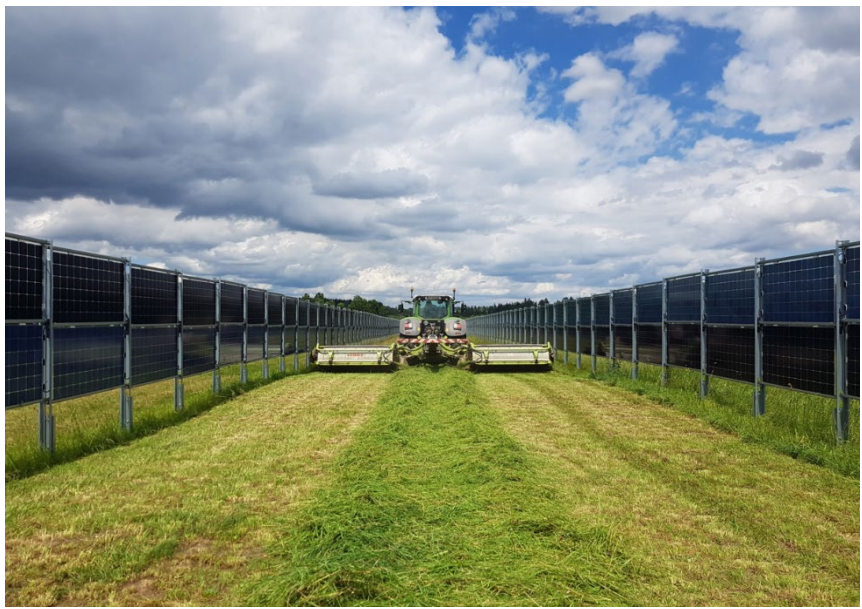

Figure 1: Vertically mounted agrivoltaic system with bifacial PV modules (Photo: Next2Sun GmbH).

# 2 Modelling and optimization

A summary of the modelling work conducted in this study is provided in Figure 2. The model calculates the PV production starting from the climatological data and applying the algorithms concerning solar position, solar decomposition and transposition, and the solar shading. Crop yield is calculated by feeding the crop yield model with PAR and other key climatological and agricultural parameters. Photosynetically active radiation (PAR) absorbed by the crop is calculated by considering the diffuse and beam components of the PAR and considering the shadings produced by the bifacial PV modules rows on the ground. The model developed in this study is based on the modelling and optimization framework of the open-source code OptiCE (OptiCE, 2021; Campana et al., 2017).

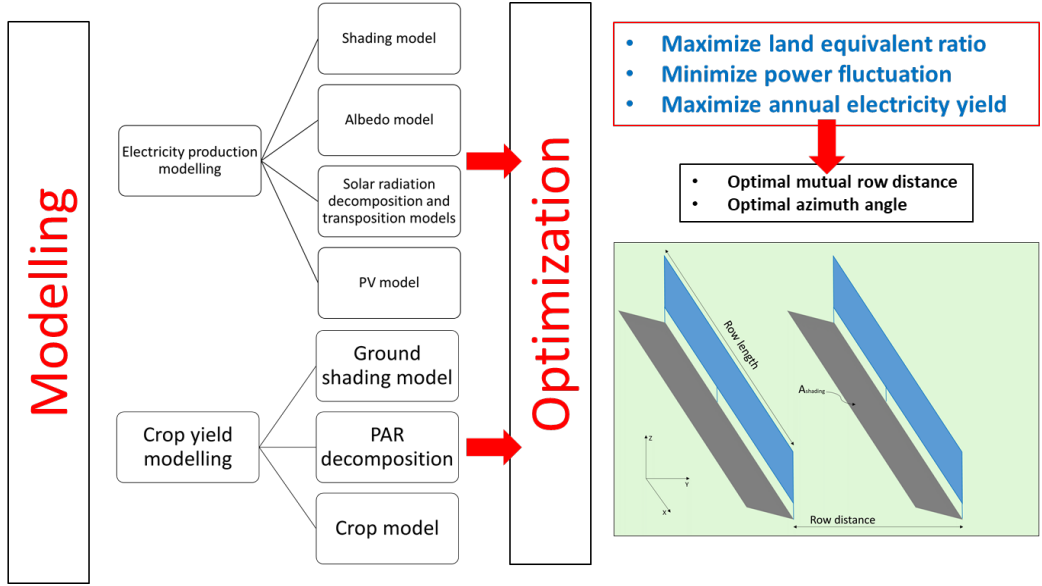

Figure 2: Modelling and optimization framework.

## 2.1 Solar radiation and photosynetically active radiation data

Data concerning meteorological variables, in particular global and diffuse horizontal solar radiation and PAR have been retrieved from the SMHI hourly data products Mesan (2021) and Strång (2021a). The climatological data refers to 2019. More information on the algorithms used by Strång can be found in Landelius et al. (2001). The data refer to a location nearby Västerås, Kärrbo Prästgård (59.5549° N, 16.7585° E) in Västmanland, where the first agrivoltaic system in Sweden will be installed.

## 2.2 Solar position algorithm

The solar position algorithm required to calculate the solar altitude $\alpha_s$ (°) and azimuth $\gamma_s$ (°) angles refers to the model developed by Meeus (1991), and Reda and Andreas (2004).

## 2.3 Solar decomposition and transposition models

The diffuse component of the global solar radiation has been retrieved from Strång (2021a) (see Section 2.1), nevertheless, due to the unavailability of the diffuse component of the PAR, the decomposition model from Gu et al. (1999) is implemented (Oliphant and Stoy, 2018). The decomposition model calculates the diffuse fraction of the global radiation. The diffuse component of the PAR has been used to calculate the total absorbed PAR by considering the effect of shadings. A validation of the PAR decomposition model is provided in Section 3. The solar transposition algorithm refers to Perez1990 (Perez et al., 1987; Perez et al., 1990) available in the *PV_LIB* (Sandia National Laboratories, 2019). The solar transposition algorithm calculates the solar radiation hitting an oriented and tilted surface starting from the global and diffuse components of the horizontal radiation. The solar radiation on the front and rear side of the bifacial PV model has been calculated with the models from Martín et al. (2005), Sun et al. (2018), and Khan et al. (2017).

## 2.4 PV model

The PV power production is calculated with the five parameters single diode model as described in Duffie and Beckman (2013). The five parameters are derived by minimizing the mean absolute error between calculated and measured I-V curve at 1000 W/m². The characteristic of the bifacial PV module refers to Jolywood D72N 380 $W_p$ (Jolywood, 2021) summarized in Table 1. The full cell N-Type module has a bifaciality of 80%.

Table 1: Bifacial PV module characteristics for the front side (Jolywood, 2021).

| Parameter | Value |
| --- | --- |
| Peak power ($W_p$) | 380 |
| MPP Voltage (V) | 40.2 |
| MPP Current (A) | 9.44 |
| Open circuit voltage (V) | 49.5 |
| Short circuit current (A) | 9.93 |
| Module efficiency (%) | 19.41 |
| Dimensions (mm) | 1974 * 992 |

We have simulated an agrivoltaic system of 22.8 $kW_p$ composed by 60 bifacial PV modules arranged in three rows. Each row is composed by 20 PV modules distributed in two rows.

## 2.5 Shading losses

The shading calculations have been performed following the approach presented in Cascone et al. (2011). The method starts from the definition of the solar vector $\vec{S}$ in a coordinate system centred on a local horizontal plane at an arbitrary location on the Earth and it is defined as follows (Stine, and Geyer, 2001):

$$\vec{S} = S_S \hat{\imath} + S_E \hat{\jmath} + S_Z \hat{k}, \tag{1}$$

where i, j, and k are unit vectors along the axis defined by the South, East, and Zenith directions, respectively (Blanco-Muriel et al., 2001). While, $S_S$, $S_E$, and $S_Z$ are defined in terms of solar altitude $\alpha_s$ (°) and azimuth $\gamma_s$ (°) angles as follows:

$$\vec{S} = \begin{cases} cos\alpha_s cos\gamma_s \\ cos\alpha_s sin\gamma_s \\ sin\alpha_s \end{cases} \tag{2}$$

The definition of $\vec{S}$ allows to calculate the projections of all the vertexes of each shading object on the horizontal plane and on the other PV planes, to calculate the shading on crops and on the PV system, respectively. The shading factor $S_F$ for each surface is then defined as follows:

$$S_F = \frac{S_{F,b} \cdot I_b + S_{F,d} \cdot I_d + I_r}{I_b + I_d + I_r}, \tag{3}$$

where, $S_{F,b}$ and $S_{F,d}$ are the shading factors for the beam and diffuse irradiance, while $I_b$, $I_d$, and $I_r$ are the direct, diffuse, and reflected irradiances on the surface without shadings. In particular, $S_{F,b}$ has been calculated as follows:

$$S_{F,b} = \frac{A_{shading}}{A_{tot}}, \tag{4}$$

where, $A_{shading}$ is the shaded area (m²) and $A_{tot}$ is the total reference area (m²). For instance, the total reference area for the crop is the area comprised between two PV modules rows as shown in Figure 3. Once the projection of P on the horizontal plane (plane were crops are grown) is calculated, $A_{shading}$ can be easily calculated through geometrical considerations (see Figure 3). In details, the model calculates the intersection between the area of the rectangles/parallelograms produced by the shadings and the total area between the rows of the bifacial PV modules using the Matlab® function *rectint*.

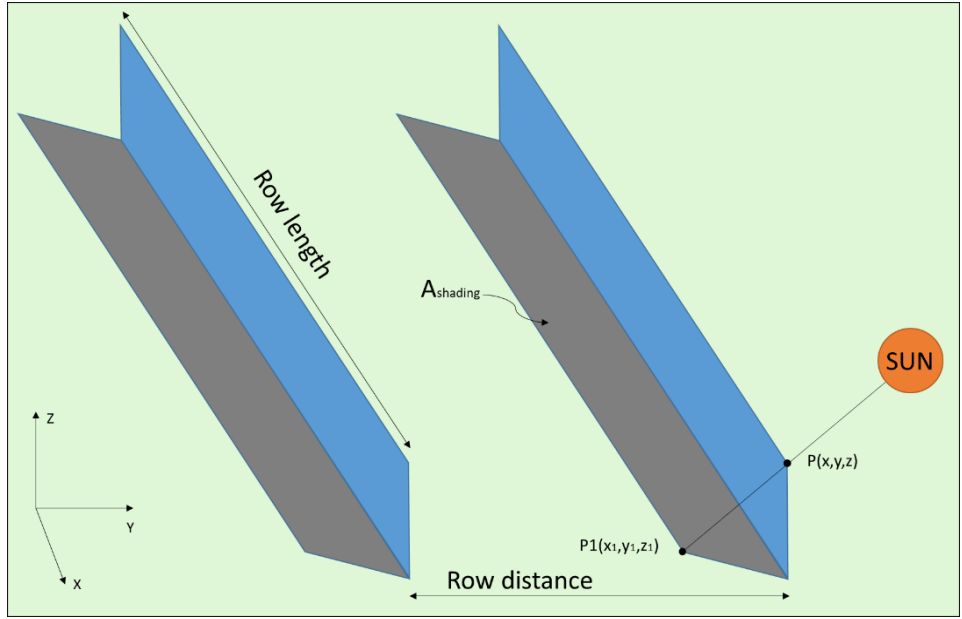

Figure 3: Calculation of the shaded area for the estimation of the shading factor.

The sky dome is discretized with altitude and azimuth angles of 1° (i.e., the mesh grid is composed by 360×90 elements). $S_{F,b}$ is imported as matrix and the values are retrieved with k-nearest neighbors algorithm similar to Melo et al. (2013). The error introduced by the k-nearest neighbors algorithm is presented in Section 3. The shading factor for the diffuse irradiance is calculated using the approach used in Li and Lam (2004) and in Cascone et al. (2011) and given by the following equation:

$$S_{F,d} = \frac{\int_{\alpha=0}^{\pi/2} \int_{\gamma=0}^{2\pi} S_{F,b} R_{\alpha\gamma} cos\theta d\Omega}{\int_{\alpha=0}^{\pi/2} \int_{\gamma=0}^{2\pi} R_{\alpha\gamma} cos\theta d\Omega}, \tag{5}$$

where, α and γ are the altitude and azimuth angles (°), $R_{\alpha\gamma}$ is the radiance (W/m²sr), θ is the angle of incidence (°), and Ω is the solid angle (sr). $R_{\alpha\gamma}$ is calculated with the modified Perez et al. (1993) model as it is in the Standard ISO 15469: 2004 (E)/CIE S 011/E: 2003 (2004) and in Darula and Kittler (2002). The horizon profile has been retrieved from PVGIS (2021) to remove those values of solar radiation and shadings when the sun position is below the horizon.

## 2.6 Crop model

The crop model refers to the Environmental Policy Integrated Climate (EPIC) model developed by Williams et al. (1989). The model has been converted into a Matlab® code by the main author and applied in Zhang et al. (2018) for corn yield modelling in Nebraska and by Campana et al. (2018) for potato yield modelling in Sweden. The model converts the daily PAR into biomass. The biomass production is then reduced due to the effects of various stress (i.e., water stress, temperature stress, or fertilizer stress) and harvest index. The harvest index is applied to calculate the economic yield of the culture. The actual crop yield $Y_a$ (t/ha) is given by the following equation (Williams et al., 1989):

$$Y_a = HIA \sum_{i=1}^{N} * BE * 0.001 * PAR_{tot} * (1 - e^{-0.65 * LAI_i}) * \gamma_{reg,i}, \tag{6}$$

where, $HIA$ is the adjusted harvest index at maturity, $BE$ is the biomass-energy ratio ((kg/ha)/(MJ/m$^2$)), $PAR_{tot}$ is the daily PAR (MJ/m$^2$), $LAI_i$ is the daily leaf area index defined as leaf area per unit ground surface area (m$^2$/m$^2$), and $\gamma_{reg,i}$ is the daily crop growth regulating factor. $\gamma_{reg,i}$ is the lowest values among the daily water, temperature, aeration, and nutrients stresses. The soil properties are also accounted in $\gamma_{reg,i}$. The reference evapotranspiration ($ET_0$) has been calculated with the Hargreaves and Samani (1985) equation as in Campana et al. (2018). $ET_0$ plays a key role in the calculation of the soil water budget and thus in the calculation of the water stress as in Allen et al. (1998). $PAR_{tot}$ has been calculated as PAR if no shading is occurring or as a contribution of PAR and $PAR_{diffuse}$ if shading is occurring. Oat and potato as reference crops to be grown between the bifacial PV modules rows. The crop model key parameters are given in Table 2 for oat and in Table 3 for potato. The validation of the crop modelling without shading effects is given in Section 3 by using field measured data and county average statistics. In particular, the crop yield validation for oat has been performed by using the 2018 agricultural data from the Lanna weather station, belonging to the Integrated Carbon Observation System (ICOS) network (ICOS, 2019; Weslien, 2020). The simulated yield has been also validated with county statistical yield (Statistic Sweden, 2018). The climatological data were retrieved from Mesan (2021) and Strång (2021a). It must be pointed out that no irrigation was considered during the simulations despite potato is one of the most irrigated crop in Sweden. An optimal nutrients management is assumed in the simulation, which means no stress is due to lack of nutrients.

Table 2: Parameters for oat.

| Parameter | Value | Reference |
|---|---|---|
| Harvest index | 0.42 | (Williams et al., 1989) |
| Biomass energy ratio ((kg/ha)/(MJ/m$^2$)) | 35 | (Williams et al., 1989) |
| Base temperature (°C) | 0 | (Williams et al., 1989) |
| Optimal temperature (°C) | 15 | (Williams et al., 1989) |
| Maximum LAI (m$^2$/m$^2$) | 5 | (Williams et al., 1989) |
| Water stress-yield factor | 0.21 | (Williams et al., 1989) |
| LAI declining factor | 1 | (Williams et al., 1989) |
| Fraction of growing season when leaf area declines | 0.8 | (Williams et al., 1989) |
| First point on optimal leaf area development curve (%) | 15.01 | (Williams et al., 1989) |
| Second point on optimal leaf area development curve (%) | 50.95 | (Williams et al., 1989) |

Table 3: Parameters for potato.

| Parameter | Value | Reference |
|---|---|---|
| Harvest index | 0.95 | (Williams et al., 1989) |
| Biomass energy ratio ((kg/ha)/(MJ/m$^2$)) | 30 | (Williams et al., 1989) |
| Base temperature (°C) | 7 | (Williams et al., 1989) |
| Optimal temperature (°C) | 20 | (Williams et al., 1989) |
| Maximum LAI (m$^2$/m$^2$) | 5 | (Williams et al., 1989) |
| Water stress-yield factor | 0.95 | (Williams et al., 1989) |
| LAI declining factor | 2 | (Williams et al., 1989) |
| Fraction of growing season when leaf area declines | 0.6 | (Williams et al., 1989) |
| First point on optimal leaf area development curve (%) | 15.01 | (Williams et al., 1989) |
| Second point on optimal leaf area development curve (%) | 50.95 | (Williams et al., 1989) |

## 2.7 Optimization

The optimization problem is a multi-objective optimization problem solved through a genetic algorithm. To analyse most of the trade-offs involved in the design and operation of agrivoltaic systems, the optimization problem maximizes the Land Equivalent Ratio (*LER*), minimizes the fluctuation of the power injected into the grid in terms of annual standard deviation (*STD*), and maximizes the annual electricity production. LER is defined as follows (Dupraz, 2011; Amaducci et al, 2018):

$$LER = \frac{Y_{c,a}}{Y_{c,r}} + \frac{E_{PV,a}}{E_{PV,r}}, \quad (7)$$

while the STD is given by Xu et al. (2013):

$$STD = \sqrt{\frac{1}{N-1}\sum_{i=1}^{N}(P_{i,t} - \overline{P_i})^2}, \quad (8)$$

where, $Y_{c,a}$ is the crop yield in the agrivoltaic configuration (t/ha), $Y_{c,r}$ is the crop yield in the reference case (i.e., open field) (t/ha), $E_{PV,a}$ is the energy conversion of the agrivoltaic system (kWh/m²/year), and $E_{PV,r}$ is the energy conversion of a conventional ground mounted PV system (kWh/m²/year), $P_{i,t}$ is the power injected into the grid (kW), and $\overline{P_i}$ is the average power injected into the grid (kW). $Y_{c,r}$ has been calculated by running the model without considering any shadings produced by the bifacial PV modules on the crop. The reference energy conversion for a ground-mounted PV system in Västmanland is assumed equal to 1000 kWh/kW$_p$ assuming a PV system tilted with 35° and a pitch of 5m as shown in Figure 4. In Yang et al. (2020), the reference system was a 35 kW$_p$ system distributed in 10 rows with 350 W$_p$ PV modules of dimensions 0.992m×1.956m. This corresponds to a system that produces 35,000 kWh/year in a net area of 600 m² (i.e., about 58 kWh/m²/year). We have assumed to connect the system directly to the grid and not cover any load, for instance farm´s load. The decisional variables of the optimization model are the azimuth angle and the mutual rows distance. The lower and upper boundaries are -180°/0°, and 5/20 m, respectively. The lower boundary for the mutual rows distance is due to the minimum space for agricultural machineries to perform conventional agricultural practices between the bifacial PV rows.

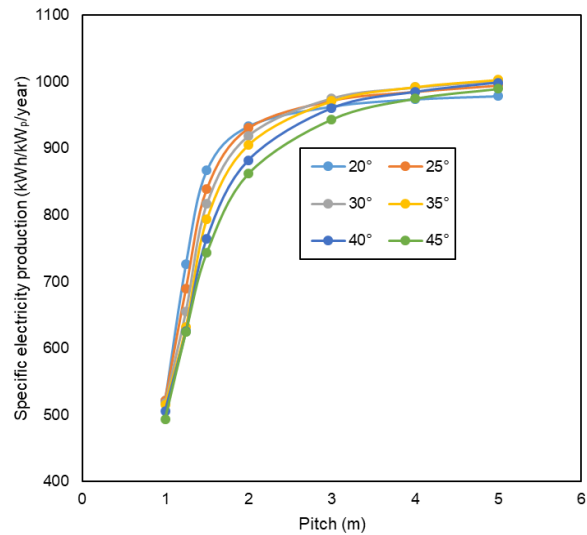

Figure 4: Specific electricity production as a function of pitch (m) and tilt angle (°) in Västerås. Adapted from Yang et al. (2020).

# 3 Results and discussion
## 3.1 Simulation and validation

The error introduced by the k-nearest neighbors algorithm in extrapolating the beam shading factor from the shading matrix is presented in Figure 5 as compared to the directly calculated beam shading factor. The coefficient of determination ($R^2$), the root-mean-square error (RMSE), mean absolute error (MAE), and mean bias error (MBE) shows that the error introduced by exacting the beam shading factor from the shading matrix is negligible as compared to a more detailed computation of the beam shading factor. This step is performed to speed the shading simulation step. A comparison between the beam shading factor calculated with the algorithm presented in 2.5 and with the commercial software PVsyst®, a reference software in the solar energy sector with one of the most powerful shading scene construction tool for simulating complex shadings affecting PV system performances, is depicted in Figure 6. As can be seen from the figure, good correlation can be found between the model developed in this study and the commercial software with an $R^2$ greater than 90%.

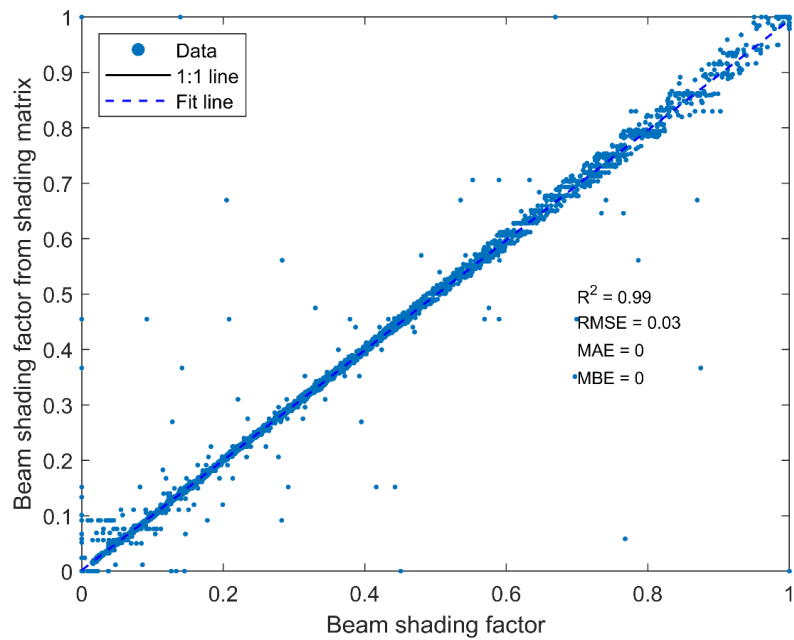

Figure 5: Comparison between beam shading factor and beam shading factor extrapolated from the shading matrix using the k-nearest neighbors algorithm (sky dome is discretized with altitude and azimuth angles of 1°).

The validation of the PAR decomposition is presented in Figure 7. Good agreement is found with the measured diffuse PAR using the data from Lanna weather station for the period 2018-2020 (Weslien, 2020). It has to be pointed out that the model developed by Gu et al. (1999) requires the calculation of the hourly diffuse horizontal transmittance calculated as the ratio between horizontal diffuse radiation and global horizontal radiation. Due to the lack of measurements at the weather station of the diffuse horizontal radiation, the diffuse horizontal radiation was calculated through the decomposition model developed by Paulescu and Blaga (2016). The decomposition step naturally introduces some errors in the validation, nevertheless, the model developed by Paulescu and Blaga (2016) showed high accuracy in Campana et al. (2020) as compared to other hourly and 1-minute decomposition models.

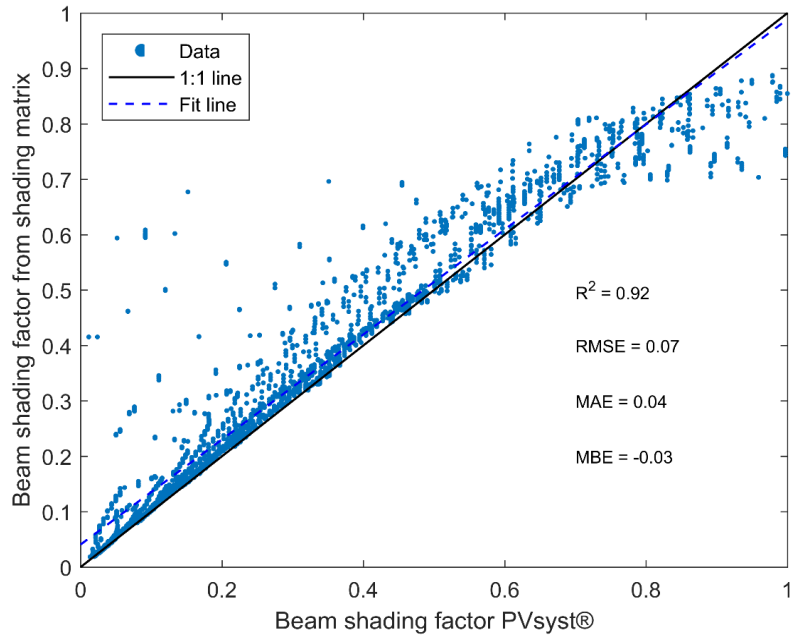

Figure 6: Comparison between beam shading factor calculated with the algorithm in section 2.5 (sky dome is discretized with altitude and azimuth angles of 1°) and with PVsyst®.

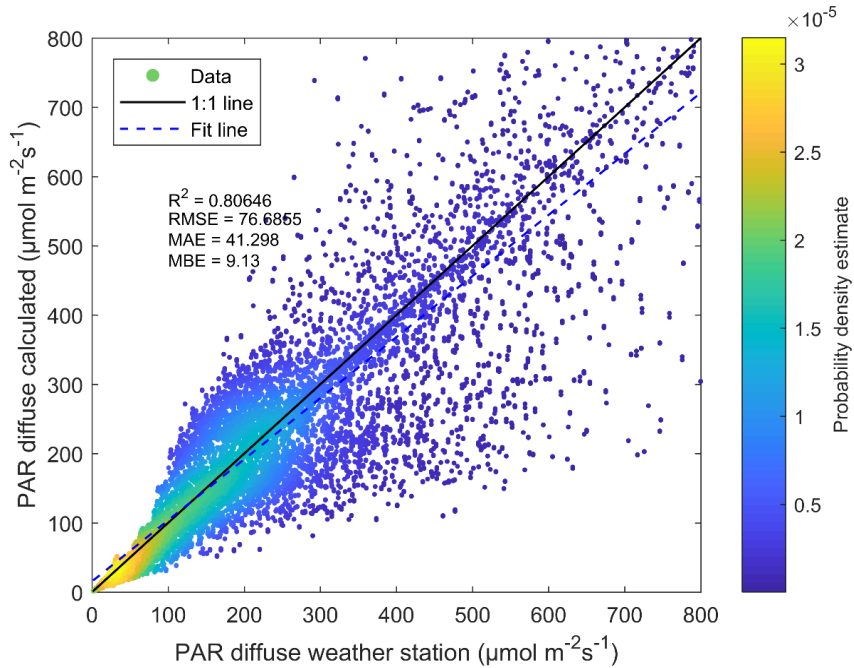

Figure 7: PAR decomposition model validation.

A straightforward sensitivity analysis showing the interrelationships between crop yield, specific electricity production, and mutual row distance is provided in Figure 8 by varying the mutual row distance between 5 and 20 m. In general, from a mutual row distance of 5 m to 20 m the crop yield doubled. The maximum oat and potato yield without shading effect were 5.1 t/ha and 9.9 t/ha, respectively. A cross validation with the commercial software PVsyst® was also performed to further analyse the error in electricity production with the model implemented in this study. In Figure 8, a good agreement between simulated value and the value obtained from the commercial software can be observed. The crop model validation in terms of specific yield (t/ha) is provided in Figure 9 using Mesan (2021) and Strång (2021a) data as climatological input data. The percentage error is 14% as compared to the actual yield measured in the farm where the Lanna weather station is located and 6% as compared to the county statistics. It has to be pointed out that the crop yield has been calculated with Mesan and Strång data and not with actual climatological data. Mesan and Strång are mesoscale models that covers the Nordic countries and the current resolution is 2.5 km. The use of input data from mesoscale models introduces an intrinsic error in the calculation of the crop yield as compared to using climatological data from weather stations. By way of example, the reader can refer to the average error statistics for hourly and daily simulations for Strång (2021b). The crop yield validation for potato was performed in Campana et al. (2018).

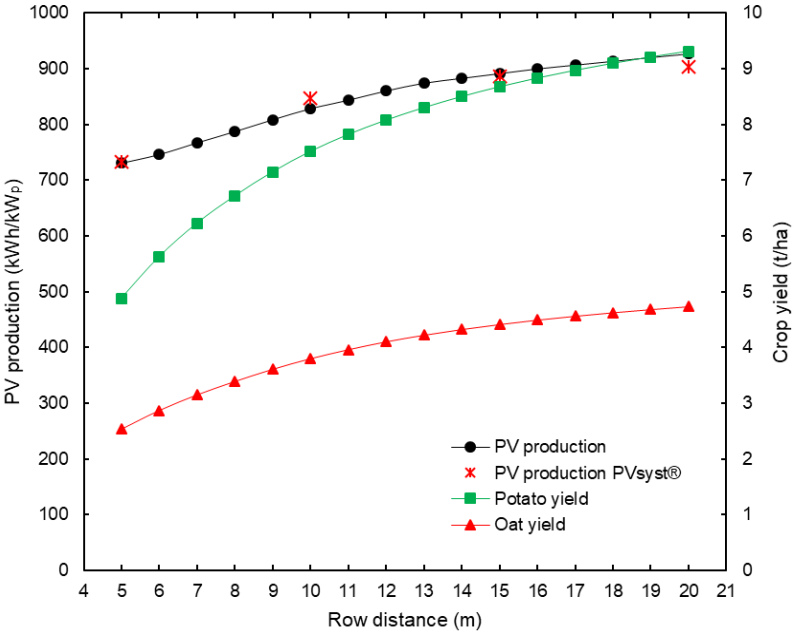

Figure 8: Relationship between row distance, specific PV production, and crop yield both for oat and potato. A validation of the specific PV system production with values obtained by the commercial software PVsyst® is superimposed.

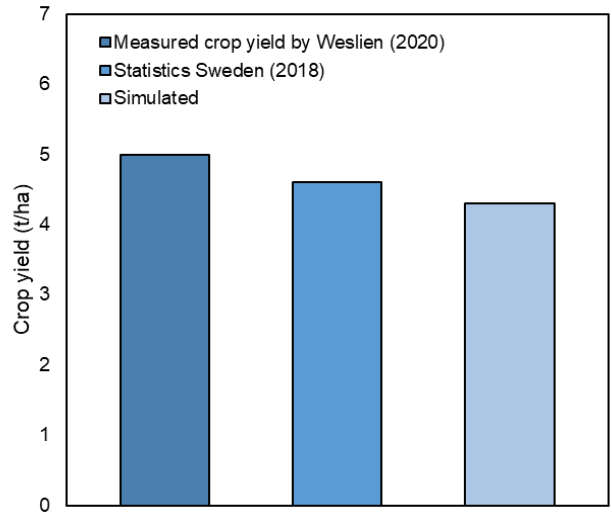

Figure 9: Validation of the crop modelling using Lanna weather station agricultural data (Weslien, 2020) and county-level statistics both for oats (Statistic Sweden, 2018).

## 3.2 Optimization

The results of the three objective optimization are depicted in Figure 10 in terms of near-optimal Pareto front and relative projections on the three surfaces (i.e., near-optimal Pareto fronts of three double-objective optimizations). The scatter plot matrices for the objectives of the optimization model and the decisional variables are depicted in Figure 11. The related Pearson coefficients between the three objectives functions and the two decisional variables are summarized in Table 4 to show which objective function drives most significantly the selection of the optimal decisional variables.

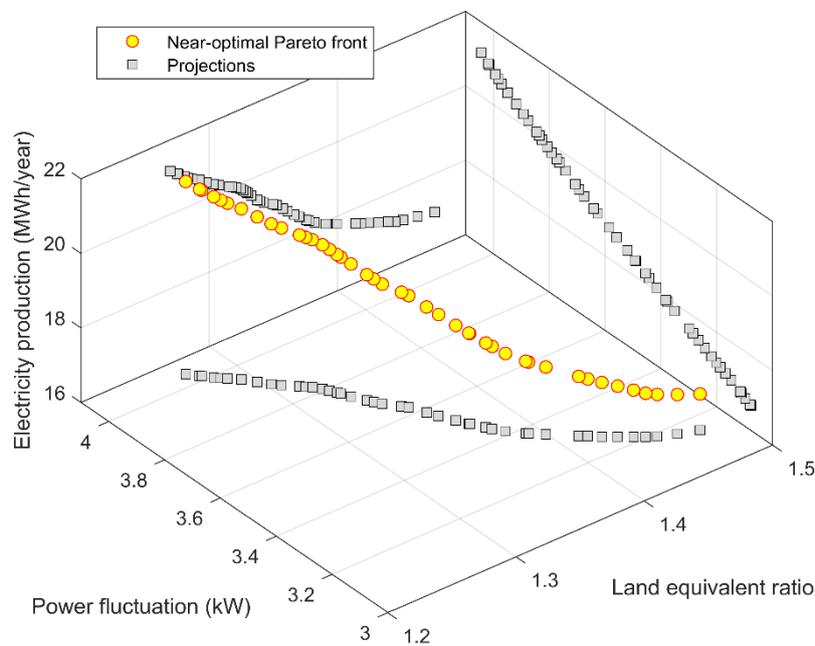

Figure 10: Near-optimal Pareto of the three objectives optimization and relative projections on the surfaces.

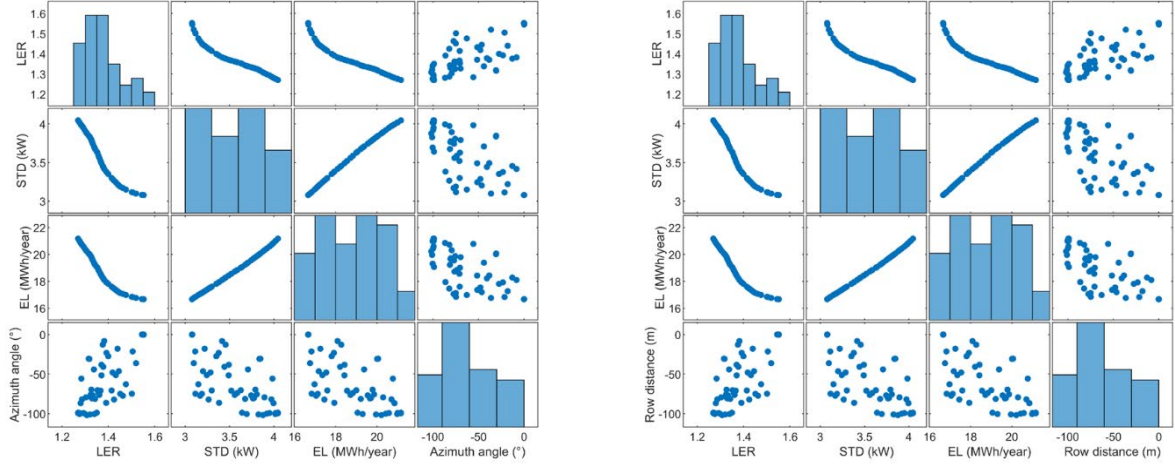

Figure 11: Scatter plot matrix of Land Equivalent Ratio (LER), power fluctuation (STD), annual electricity production (EL), and azimuth angle (left) and row distance (right).

Table 4: Pearson correlation coefficients between the three objectives and the decisional variables.

| Objective function | Decisional variables | |
|---|---|---|
| | Azimuth angle (°) | Row distance (m) |
| **max LER** | **0.451** | -0.8461 |
| **min STD** | -0.244 | 0.9704 |
| **max annual production** | -0.2221 | **0.9802** |

Electricity production and power fluctuation decrease significantly while increasing the *LER* (Figure 10). Obviously, the STD and the annual production show a linear correlation since by increasing the PV power production the power fluctuation in the grid increases. Row distance has a negative correlation coefficient (-0.8461) with the first objective (i.e., maximizing the *LER*). By increasing the mutual rows distance, the *LER* decreases. Despite a lower distance between the mutual rows reduces the crop yield and the PV production, *LER* increases since the specific PV production per unit of area increases and its contribution is more significant than the crop contribution to the *LER*. This can be clearly seen in Figure 12 where the *LER* for potato is split in the two contributions (i.e., the *LER* related to the crop yield and to the electricity production). It is interesting to note that the row distance of 9 m configures as an optimal trade-off for oat (left) while 8.5 m for potato (right). These results have a fundamental effect on the design of agrivoltaic systems since typically crop rotation is performed by farmers. It is important to note that the implementation of agrivoltaic system can lead to LER above 1.2, which means a 20% more productivity for the same land as compared to monoculture. Values above 1.5 were reported by Trommsdorff et al. (2021) in Germany, while values above 1.2 were reported by Amaducci et al. (2018) in Italy. It must be pointed out that different crops, location and thus solar radiation distribution, and agrivoltaic configurations can lead to different performances. The second objective (i.e., minimizing the power fluctuation) drives the selection of azimuth angle towards east since the electricity production profile shows lower peaks as compared to PV modules oriented towards south as depicted in Figure 13. The third objective tends to increase the mutual distance between the rows of the bifacial PV modules and it selects optimal azimuth angles towards the south. Indeed, azimuth angles towards the south lead to an increase of the electricity production as compared to azimuth angles towards east (i.e., PV modules rows in the east-west orientation). The scatter plot with density concerning the optimal azimuth and row distance as a function of the annual electricity production are depicted in Figure 143. Optimal azimuth angles are around -40°. Indeed, the electricity production of a PV system oriented at -90° and at -40° is 19.6 MWh/year and 20.1 MWh/year, respectively. Azimuth angle is mostly affected by the first objective (positive correlation) with Pearson correlation coefficient equals to 0.451 and by the second objective (negative correlation) with Person correlation coefficient equals to -0.244 (Table 4). The strong correlation between the azimuth angle and the objective of maximizing the LER can also be seen in Figure 11(left) (upper-right plot). The choice of the optimal row distance is mainly driven by the third objective with a positive correlation (i.e., Pearson correlation coefficient of 0.9802).

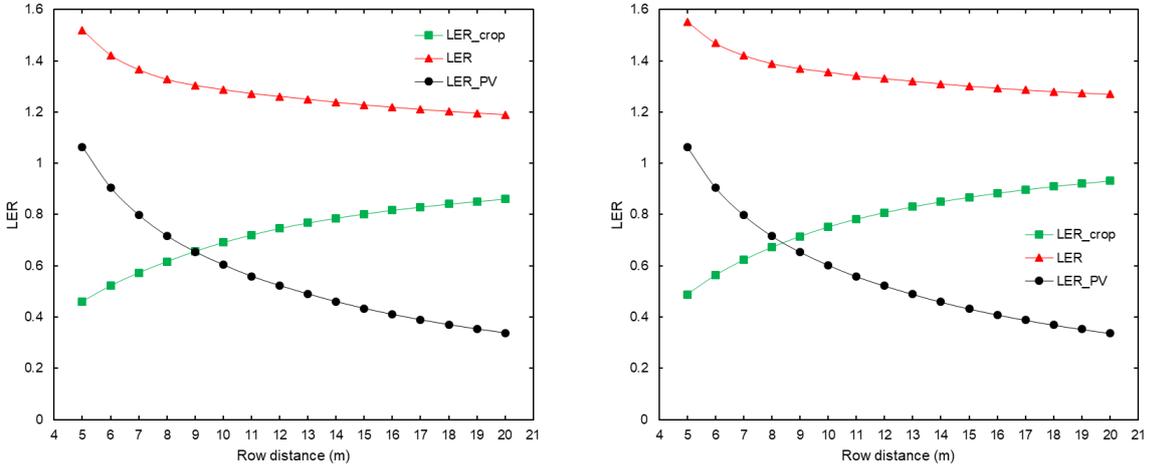

Figure 12: Land Equivalent Ratio (*LER*) and its contributions (i.e., the *LER* related to the crop yield and to the electricity production) for oat (left) and potato (right).

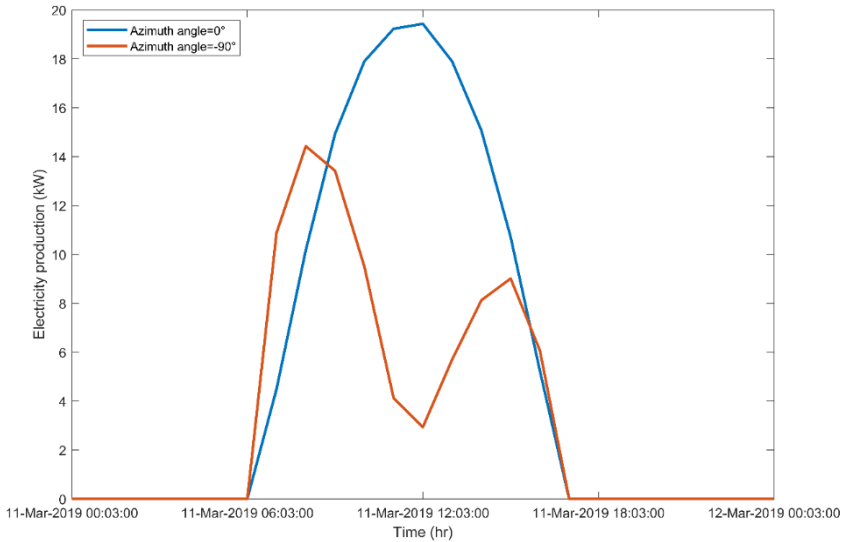

Figure 13: Hourly electricity profile for a PV system oriented towards 45° east and towards east.

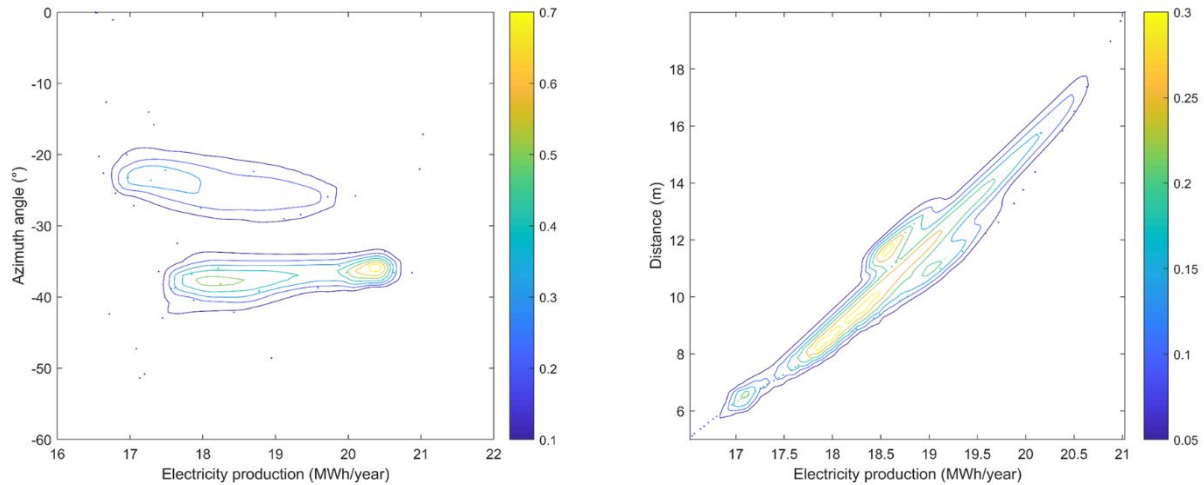

Figure 14: Optimal azimuth angles and distances distributions as compared to the annual electricity production (objective 3).

One of the limitation of this work is that the optimization should be carried out considering several years and different crops rotations as highlighted in Figure 12. Nevertheless, this is beyond the scope of this study, which focused mostly on the model development. Multi-year and multi-crop optimizations will be undertaken in future studies. Moreover, more in details economic analysis based on comprehensive and lifetime-oriented indicators, such as the net present value (NPV), will be carried out in future studies. Most of the objective functions used in this work are technical oriented, more economic as well as environmental oriented objective functions will be implemented in the future version of the model.

## 4 Conclusions

In this study, an optimization algorithm for vertically mounted agrivoltaic systems using bifacial PV modules has been developed and applied to two different crops for a location in Sweden. To the best knowledge of the authors this is one of the first attempts in optimizing agrivoltaic systems. The main conclusions drawn from this study are the following:

- The agrivoltaic model presented in this study is composed by several sub-models. In particular, the models concerning shadings, photosyntetically active radiation decomposition, crop yield, and PV production model are the most fundamental. The validation of those sub-models shows the robustness of the model to identify the interrelationships between crop productivity and electricity production based on the system configuration (i.e., orientation and row distance).
- The row distance affects significantly the photosyntetically active radiation distribution and at the latitude of the investigated location, crop yield can be halved by passing from 20 m to 5 m distance between mutual rows.
- By analysing the optimization results, it is clear that the Land Equivalent Ratio cannot be used as main parameter for the optimal design of agrivoltaic systems. More objective functions should be included for a better estimation of the synergies between crop and electricity production. Maximizing the Land Equivalent Ratio tends to drastically reduce electricity production and this could undermine the investment on the PV system.
- By studying the contributions of Land Equivalent Ratio (i.e., the crop yield and the electricity production contributions), it is evident that the optimal row distance varies according to the crop

(i.e., 8.5 m for potato and 9.0 m for oat) and this leads to important consequences in terms of long term optimal design of an agrivoltaic system. An optimal design of the agrivoltaic system should consider multi-year and multi crop simulations and optimization based on conventional farm activities.
- The implementation of agrivoltaic system for the investigated crops and locations shows land equivalent ratios above 1.2, as also estimated in previous studies conducted in Germany and Italy. These results justify the implementation of agrivoltaic systems to meet green economy and sustainability goals.

## Acknowledgements

The main author and Bengt Stridh acknowledges the Swedish Energy Agency for the funding received through the project "Evaluation of the first agrivoltaic system in Sweden". The authors also acknowledge ICOS for providing the data gathered at Lanna station used in this study to validate the crop model.